\documentclass[twocolumn,showpacs,preprintnumbers,amsmath,amssymb,prl]{revtex4}
\usepackage{graphicx}% Include figure files
\usepackage{dcolumn}% Align table columns on decimal point
\usepackage{bm}% bold math

\begin{document}

\title{Orthorhombic-to-Monoclinic Phase Transition of Ta$_2$NiSe$_5$ Induced by the Bose-Einstein Condensation of Excitons}

\author{T. Kaneko,$^1$ T. Toriyama,$^1$ T. Konishi,$^2$ and Y. Ohta$^1$}
\affiliation{$^1$Department of Physics, Chiba University, Chiba 263-8522, Japan}
\affiliation{$^2$Graduate School of Advanced Integration Science, Chiba University, Chiba 263-8522, Japan}

%\date{\today}
\date{October 8, 2012}

\begin{abstract}
Using the band structure calculation and mean-field analysis of the 
derived three-chain Hubbard model with phonon degrees of freedom, 
we discuss the origin of the orthorhombic-to-monoclinic phase transition 
of the layered chalcogenide Ta$_2$NiSe$_5$.  
We show that the Bose-Einstein condensation of excitonic electron-hole 
pairs cooperatively induces the instability of the phonon mode at momentum 
$q\rightarrow 0$ in the quasi-one-dimensional Ta-NiSe-Ta chain, resulting in the 
structural phase transition of the system.  The calculated single-particle 
spectra reproduce the deformation of the band structure observed in the 
angle-resolved photoemission spectroscopy experiment.  
\end{abstract}


\maketitle

The electron-hole pair condensation in thermal equilibrium into the 
excitonic insulator (EI) state has attracted renewed interest in 
recent years because of the discoveries of a number of new materials.  
The idea of EI was proposed about half a century ago \cite{mott,knox}, 
where the EI state was predicted to be realized either in a 
semiconductor with a small band gap or in a semimetal with a small 
band overlap.  The formation of excitons is driven by poorly screened 
Coulomb interaction between conduction-band electrons and 
valence-band holes under condition of a low carrier concentration.  
If the binding energy of excitons is larger than the band gap, they 
may spontaneously condense at low temperatures and drive the system 
into a new ground state with exotic properties.  This new state, 
which is a condensed state of a macroscopic number of excitons 
acquiring quantum phase coherence, is called EI \cite{jerome,halperin}.  
It has been pointed out that the semimetal-EI transition may be 
described in analogy with the BCS theory of superconductivity and 
the semiconductor-EI transition is discussed in terms of a 
Bose-Einstein condensation (BEC) of preformed excitons 
\cite{bronold,ihle,phan,seki,zenker}.  

An example of materials for possible realization of EI is Tm(Se,Te), 
where it has been claimed that a transition into EI occurs by 
applying pressure \cite{bucher,wachter1,wachter2} and that the 
superfluidity of condensed excitons is responsible for the observed 
anomalous properties \cite{bronold2}.  
Another promising candidate for EI is CaB$_6$ \cite{young}, where 
the observed weak ferromagnetism with an unexpectedly high Curie 
temperature was interpreted in terms of a doped EI 
\cite{zhitomirsky,bascones}.  
Also known as a candidate for EI is TiSe$_{2}$, where the observed 
charge-density-wave state was interpreted to be of an excitonic type 
\cite{cercellier,wezel,monney}.  
The spin-density-wave state of the iron-pnictide superconductors 
has also been argued to be of the excitonic type 
\cite{brydon,zocher,kaneko}.  
Semiconductor bilayer systems have attracted attention as well 
in relation to the BEC of excitons \cite{eisenstein}.  

Recently, a transition-metal chalcogenide Ta$_{2}$NiSe$_{5}$ has been 
studied in this respect \cite{wakisaka1,wakisaka2}.  This material has 
a layered structure stacked loosely by a weak van der Waals interaction, 
and in each layer, Ni single chains and Ta double chains are running 
along the $a$-axis of the lattice to form a quasi-one-dimensional (1D) 
chain structure \cite{sunshine}.  
The observed resistivity shows a semiconducting behavior over a wide 
temperature range with a quasi-1D anisotropic electron conduction at 
high temperatures \cite{disalvo}.  Then, an anomaly in the resistivity 
appears at 328K, which is associated with a second-order-like 
structural phase transition from orthorhombic to monoclinic phase 
\cite{disalvo}.  
The magnetic susceptibility exhibits diamagnetism in a wide temperature 
range ($4.2-900$ K) and shows a sudden drop (being more negative) 
below the structural transition temperature (328 K) \cite{disalvo}.  
The system was thus suggested to be a small band-gap semiconductor with 
oxidation states of Ni$^{0+}$ ($3d^{10}$) and Ta$^{5+}$ ($5d^0$), rather 
than a magnetic or Mott insulator \cite{disalvo,canadell}.  
However, a recent X-ray photoemission spectroscopy (XPS) experiment,  
together with a cluster-model calculation, showed that Ni ions have a 
$3d^{9}\underline L$ character ($\underline L$ is a Se 4$p$ hole) 
and consequently Ta ions have a $5d^1$ character \cite{wakisaka1}.  
Moreover, the angle-resolved photoemission spectroscopy (ARPES) 
experiment \cite{wakisaka1,wakisaka2} showed that the spectra are 
strongly temperature dependent, i.e., at 40 K the flatness of the top 
of the valence band is extremely enhanced and the size of the band 
gap becomes wider.  It was thereby suggested that the EI state is 
realized as the ground state of this material, where the spin-singlet 
excitons between the Ni 3$d$-Se 4$p$ holes and Ta 5$d$ electrons 
are presumed \cite{wakisaka1}.  

In this paper, motivated by such developments in the field, we make 
a theory to elucidate the origin of the structural phase transition 
and associated anomalous electronic properties of Ta$_2$NiSe$_5$.  
We first carry out the density-functional-theory (DFT) based 
electronic structure calculations for the orthorhombic phase of 
Ta$_2$NiSe$_5$ and find that the system is a direct-gap semiconductor 
with the gap minimum at the $\Gamma$-point of the Brillouin zone and 
has a simple band structure near the Fermi level.  
No hybridization occurs between the top of the valence band and 
bottom of the conduction band.  
Based on the results, we construct an effective three-chain Hubbard 
model to reproduce the three bands near the Fermi level.  
We then introduce the phonon degrees of freedom into the model and 
analyze it by the mean-field approximation.  We calculate the 
ground-state and finite-temperature phase diagrams of the model to 
clarify the origin of the structural phase transition.  
The single-particle excitation spectra are also calculated.  

We thus show that the BEC of excitonic electron-hole pairs cooperatively 
induces the instability of the phonon mode at momentum $q\rightarrow 0$ 
in the quasi-1D Ta-NiSe-Ta chain, resulting in the structural phase 
transition of the system.  
The spontaneous hybridization between the conduction and valence 
bands well explains the valence states of Ni and Ta ions observed 
in the XPS experiment \cite{wakisaka1} and also the calculated 
single-particle spectra well reproduce the deformation in the band 
structure near the Fermi level observed in the ARPES experiment 
\cite{wakisaka1,wakisaka2}.  
Our results thus demonstrate a possible realization of the EI state 
in the simplest quasi-1D electronic system of the recently 
corroborated material Ta$_2$NiSe$_5$ and will encourage further 
experimental studies of this intriguing material.  

%%%% FIG.1 %%%%
\begin{figure}[tbh]
\begin{center}
\includegraphics[width=6.0cm]{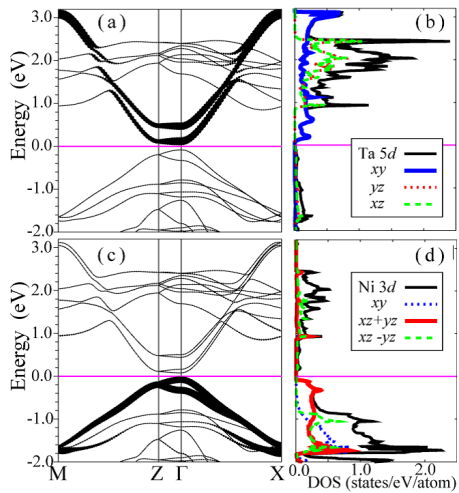}
\caption{(Color online) Calculated band dispersions and PDOS of 
the orthorhombic phase of Ta$_2$NiSe$_5$.  
In (a), the width of the curves is in proportion to the weight 
of the Ta $5d_{xy}$ orbital, in (b), the contribution from the 
Ta $5d$ $t_{2g}$ orbitals is shown, in (c), the the width of 
the curves is in proportion to the weight of the Ni $3d_{xz+yz}$ 
orbital, and in (d), the contribution from the Ni $3d$ $t_2$ 
orbitals is shown.  The Fermi level is indicated by a horizontal 
line.  The local coordinate system defined in 
Ref.~\onlinecite{wakisaka1} is used.  
}\label{fig1}
\end{center}
\end{figure}
%%%%%%%%%%%%%%%

We make the band structure calculations employing the code WIEN2k 
\cite{wien2k} based on the full-potential linearized 
augmented-plane-wave method, where we use the generalized gradient 
approximation for electron correlations with the exchange-correlation 
potential of Ref.~\onlinecite{PBE96}.  
Because the crystal structure of the high-temperature orthorhombic 
phase of Ta$_2$NiSe$_5$ is not known, we make the structural 
optimization, keeping the orthorhombic structure (space group 
$Cmcm$) and lattice constants \cite{sunshine}, and obtain the 
optimized internal coordinates of all the ions \cite{SM}.  
The calculated results for the band structure, however, predict a 
metallic state with a small band overlap and are not consistent 
with experiment.  This may be due to a well-known problem of DFT-based 
band calculations, where the band gap in semiconductors is 
underestimated.  To amend this, we shift the conduction (valence) 
bands upward (downwards) by adding (subtracting) orbital-dependent 
potentials into the Hamiltonian as a standard procedure \cite{park,lin} 
and reproduce the experimental band gap \cite{SM}.  

The results for the partial densities of states (PDOS) and band 
dispersions thus obtained are shown in Fig.~\ref{fig1}.  We find 
that the system is a direct-gap semiconductor with the gap minimum 
at the $\Gamma$-point of the Brillouin zone in agreement with 
experiment \cite{wakisaka1,disalvo}.  The band structure near the 
Fermi level is rather simple; the conduction band has a cosine-like 
quasi-1D band dispersion coming from the $5d_{xy}$ orbitals of Ta 
ions arranged along the chain, whereas the top of the valence band 
has a quasi-1D dispersion coming from the Ni $3d_{xz+yz}$ and Se 
$4p_{x+y}$ orbitals arranged along the chain, and no hybridization 
occurs between the top of the valence band and bottom of the 
conduction band.  

%%%% FIG.2 %%%%
\begin{figure}[tbh]
\begin{center}
\includegraphics[width=7.6cm]{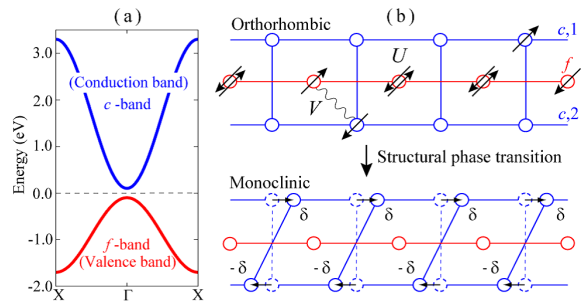}
\caption{(Color online) (a) Noninteracting band dispersion of our 
model Eq.~(\ref{ham}).  The conduction band is doubly degenerate.  
(b) Schematic representations of the three-chain Hubbard model for 
the 1D structural unit consisting of the two chains of the Ta 
$5d_{xy}$ orbitals yielding the $c$-bands and one chain of the Ni 
$3d_{xz+yz}$ and Se $4p_{x+y}$ orbitals of Ta$_{2}$NiSe$_{5}$ 
yielding the $f$-band.  The lattice distortion of the chain 
corresponding to the orthorhombic-to-monoclinic phase transition 
is schematically illustrated.  
}\label{fig2}
\end{center}
\end{figure}
%%%%%%%%%%%%%%%

Based on the results, we make the effective three-chain model 
containing the valence $f$-band coming from the hybridized Ni 
$3d_{xz+yz}$ and Se $4p_{x+y}$ orbitals and the doubly degenerate 
conduction $c$-bands coming from the Ta $5d_{xy}$ orbitals, 
to which the Hubbard-type onsite ($U_c$ and $U_f$) and interchain 
($V$) repulsive interactions are added (see Fig.~\ref{fig2}).  
We also introduce the electron-phonon interaction with strength 
$g$ and the phonon term with phonon frequency $\omega_c$ assuming 
a uniform shear distortion of the chain (see Fig.~\ref{fig2}) 
corresponding to the structural transition from the orthorhombic 
to monoclinic phase \cite{ssh,SM}.  
The Hamiltonian reads 
\begin{eqnarray}
\mathcal{H}&=&\sum_{k,\sigma,\alpha}\varepsilon_{c}(k)c_{k,\sigma,\alpha}^{\dag}c_{k,\sigma,\alpha} 
+\sum_{k,\sigma}\varepsilon_{f}(k)f_{k,\sigma}^{\dag}f_{k,\sigma} \notag \\
&+&U_c\sum_{i,\sigma,\alpha}n^{c}_{i,\uparrow,\alpha}n^{c}_{i,\downarrow,\alpha} 
+U_f\sum_{i,\sigma}n^{f}_{i,\uparrow}n^{f}_{i,\downarrow} \notag \\
&+&V\sum_{i,\sigma,\sigma',\alpha}(n^{c}_{i,\sigma,\alpha} 
+n^{c}_{i+1,\sigma,\alpha} )n^{f}_{i,\sigma'} \notag \\
&+&\frac{g}{\sqrt{L}}\sum_{k,q,\sigma,\alpha}\Big[(b_{q,\alpha}+b^{\dag}_{-q,\alpha}) 
c^{\dag}_{k+q,\sigma,\alpha} f_{k,\sigma} + {\rm H.c.} \Big] \notag \\
&+&\omega_c\sum_{q,\alpha}b^{\dag}_{q,\alpha}b_{q,\alpha}
\label{ham}
\end{eqnarray}
where $c^{\dag}_{k,\sigma,\alpha}$ and $f^{\dag}_{k,\sigma}$ are 
the Fourier transforms of $c^{\dag}_{i,\sigma,\alpha}$ that creates 
an electron with spin $\sigma$ $(=\uparrow,\downarrow)$ at site $i$ 
on the $c$-orbital of the chain $\alpha$ $(=1,2)$ and 
$f^{\dag}_{i,\sigma}$ that creates an electron at site $i$ of the 
$f$-orbital, respectively.  
$n^{c}_{i,\sigma,\alpha}$ and $n^{f}_{i,\sigma}$ are the 
electron number operators on the $c$- and $f$-orbitals, respectively.  
The noninteracting band dispersions are given as 
$\varepsilon_{c}(k)=2t_{c}(\cos k-1)+D/2-\mu$ and 
$\varepsilon_{f}(k)=2t_{f}(\cos k-1)-D/2-\mu$ with 
the hopping parameters $t_{c}$ and $t_{f}$ and the band gap $D$.  
We use the values $t_c=-0.8$, $t_f=0.4$, and $D=0.2$ in units of eV 
obtained from the fitting to the calculated band dispersion 
(see Fig.~\ref{fig1}).  $\mu$ is the chemical potential.  
We assume $U_c=U_f$ $(=U)$ for simplicity and choose $V=U/4$.  
The bosonic operator $b^{\dag}_{q,\alpha}$ creates a phonon with 
momentum $q$ in the chain $\alpha$.  
$L$ is the number of the unit cell, where the unit cell contains 
an $f$- and two $c$-orbitals.  We restrict ourselves to the filling 
of two electrons per unit cell.  

We apply the mean-field approximation 
$n_{i,\uparrow}n_{i,\downarrow}
\rightarrow\langle n_{i,\uparrow}\rangle n_{i,\downarrow}
+n_{\uparrow}\langle n_{i,\downarrow}\rangle$ 
for the onsite terms of both $c$- and $f$-orbitals and 
$n^c_{i,\sigma,\alpha}n^f_{j,\sigma}
\rightarrow\langle n^c_{i,\sigma,\alpha}\rangle n^f_{j,\sigma}
+n^c_{i,\sigma,\alpha}\langle n^f_{j,\sigma}\rangle
-\langle f_{j,\sigma}^{\dag}c_{i,\sigma,\alpha}\rangle c_{i,\sigma,\alpha}^{\dag}f_{j,\sigma} 
-f_{j,\sigma}^{\dag}c_{i,\sigma,\alpha} \langle c_{i,\sigma,\alpha}^{\dag}f_{i,\sigma}\rangle$
for the interchain term.  
We define the order parameter for the spin-singlet excitons as 
$\Delta_{\alpha}=|\Delta|e^{i\theta_{\alpha}}=\frac{V}{2L}\sum_{k,\sigma}
\langle c_{k,\sigma,\alpha}^{\dag}f_{k,\sigma}\rangle$, excluding 
possibility of the formation of any density waves because the top 
of the valence band and bottom of the conduction band are both at the 
$\Gamma$-point of the Brillouin zone \cite{halperin}.  Accordingly, the 
order parameter for the uniform lattice distortion is defined as 
$\delta_{\alpha}=2g\langle b_{q=0,\alpha}\rangle/\sqrt{L}$ 
with $\langle b_{q,\alpha}\rangle\ne 0$ only at $q=0$, 
whereby the shear distortion of the chain corresponding to the structural 
transition from the orthorhombic to monoclinic phase can be described by 
$|\delta|$ with $\delta_1=|\delta|$ and $\delta_2=-|\delta|$.  
We use a parameter $\lambda=g^2/\omega_c$ in the following discussions.  
We define $n$ to be $n=\langle n^{c}_{i,\sigma,\alpha}\rangle$ 
and $1-2n=\langle n^{f}_{i,\sigma}\rangle$, which is a measure of the 
spontaneous $c$-$f$ hybridization at $T=0$ K.  
Following the standard procedure of the mean-field theory \cite{SM}, 
we derive the mean-field Hamiltonian and diagonalize it by the Bogoliubov 
transformation.  The self-consistent (or gap) equations are derived by the 
minimization of the free energy, which are solved numerically to obtain 
the temperature dependence of $n$, $|\delta|$, and $|\Delta|$.  

%%%% FIG.3 %%%%
\begin{figure}[thb]
\begin{center}
\includegraphics[width=8.0cm]{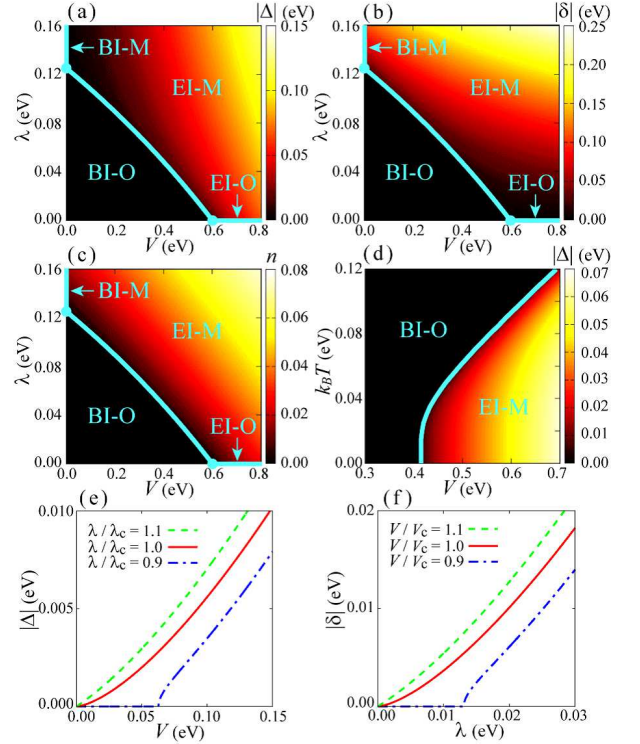}
\caption{(Color online) Ground-state phase diagram with the order parameters 
(a) $|\Delta|$, (b) $|\delta|$, and (c) $n$, in the parameter space $(V,\lambda)$.  
(d) Finite-temperature phase diagram with the order parameter $|\Delta|$ as 
a function of $V$ at $\lambda=0.04$ eV.  
Also shown are (e) the $V$ dependence of $|\Delta|$ and (f) $\lambda$ dependence of 
$|\delta|$ at $T=0$ K with $\lambda_c=0.125$ eV and $V_c=0.6$ eV.  
Abbreviations are BI (band insulator), EI (excitonic insulator), O (orthorhombic), 
and M (monoclinic).  
}\label{fig3}
\end{center}
\end{figure}
%%%%%%%%%%%%%%%

The calculated results for the ground-state and finite-temperature 
phase diagrams are shown in Fig.~\ref{fig3}.  As for the ground state, 
we clearly see in Figs.~\ref{fig3} (a)-(c) that when $V$ and $\lambda$ are 
small the system is the band insulator (BI) with $|\Delta|=|\delta|=0$ 
but when $V$ and $\lambda$ are large the EI state $|\Delta|>0$ with the 
shear distortion of the chain $|\delta|>0$ appears in the ground 
state.  The EI state occurs simultaneously and cooperatively with 
the lattice distortion except at the lines $\lambda=0$ and $V=0$.  
We emphasize here that even if $\lambda$ is small the structural phase 
transition occurs with help of the exciton condensation: i.e., the 
interaction $V$ drives the EI-state formation $\Delta>0$, which leads 
to the spontaneous $c$-$f$ hybridization, and as a consequence, the 
structural distortion $\delta>0$ occurs even if $\lambda$ is small.  
The spontaneous $c$-$f$ hybridization stabilizes the lattice distortion, 
leading to the orthorhombic-to-monoclinic phase transition.
This is consistent with the situation in the real material Ta$_2$NiSe$_5$, 
where the monoclinic distortion of the angle $0.5^\circ-1^\circ$ (or the 
atomic displacement $0.02-0.04$ \AA) is very small \cite{sunshine,disalvo}; 
a rough estimation may be $|\delta|\sim 0.02-0.04$ eV, which correspons 
to $\lambda\sim 0.02-0.05$ eV \cite{SM}.  
The oxidation states of Ni$^{2+}$ and Ta$^{4+}$ observed in the XPS 
experiment \cite{wakisaka1} are also consistent with the nonzero 
value of $n$ induced by the spontaneous $c$-$f$ hybridization in 
the EI phase (see Fig.~\ref{fig3}(c)).  

The temperature dependence of the EI phase is given in 
Fig.~\ref{fig3}(d), where we find that by lowering temperature 
the BI state with the undistorted (orthorhombic) structure changes 
into the EI state with the distorted (monoclinic) structure and the 
transition is of the second order.  Thus, the experimental situations 
are correctly reproduced.  We also point out that the observed sudden 
drop in the uniform magnetic susceptibility at the transition 
\cite{disalvo} may be due to the formation of spin-singlet 
excitons that suppresses the paramagnetic part of the uniform 
spin susceptibility of thermally activated electrons.  

%%%% FIG.4 %%%%
\begin{figure}[thb]
\begin{center}
\includegraphics[width=8.0cm]{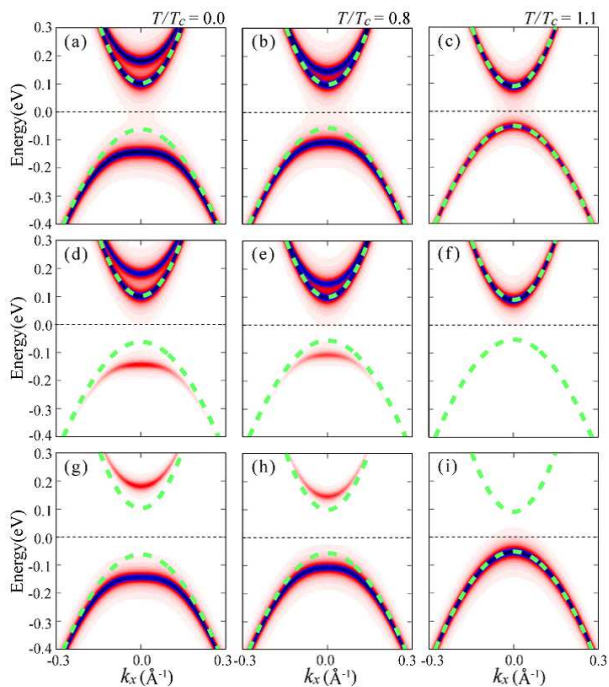}
\caption{(Color online) Calculated temperature dependence of the 
single-particle spectra of the three-chain Hubbard model.  
(a)-(c): Total spectral weight, 
(d)-(f): spectral weight on the $c$-orbital, and 
(g)-(i): spectral weight on the $f$-orbital.  
We assume $V=0.55$ eV and $\lambda=0.04$ eV.  
The calculated transition temperature is $k_{\rm B}T_c=0.0783$ eV.  
The Lorentzian broadening of the spectra of $0.01$ eV is applied for 
comparison with the experimental ARPES spectra \cite{wakisaka1,wakisaka2}.  
The band dispersion without condensation $|\Delta|=|\delta|=0$ is given 
by the dashed curves.  The Fermi level is indicated by a horizontal line.  
}\label{fig4}
\end{center}
\end{figure}
%%%%%%%%%%%%%%%

The calculated temperature dependence of the single-particle spectra 
is shown in Fig.~\ref{fig4}, where we find the large deformation of 
the top of the valence band caused by the spontaneous $c$-$f$ 
hybridization in the EI state formation by lowering temperature: 
the flattening of the band dispersion is evident and the shift of 
the top of the valence band away from the Fermi level is noticed, 
both of which are consistent with experimental ARPES results 
\cite{wakisaka1,wakisaka2}.  The hybridization can be seen in the 
$c$-orbital spectral weight {\em below} the Fermi level 
(see Figs.~\ref{fig4} (d) and (e)) and $f$-orbital spectral weight 
{\em above} above the Fermi level (see Figs.~\ref{fig4} (g) and 
(h)).  Splitting of the $c$-bands into two is also noticed, where 
the lower ``nonbonding'' band remains unaffected in the presence 
of $|\Delta|$ and $|\delta|$.  
Note that the $c$-$f$ hybridization is absent above the transition 
temperature ($T_c$) in our mean-field calculations (see Figs.~\ref{fig4} 
(c), (f), and (i)).  However, the characteristic temperature scale 
associated with the formation of preformed excitons should be 
present \cite{bronold}, which we hope will be observed in any 
future experiment.  

In summary, we have discussed the origin of the orthorhombic-to-monoclinic 
phase transition of the layered dichalcogenide Ta$_2$NiSe$_5$ using the 
band structure calculation and mean-field analysis of the derived 
three-chain Hubbard model.  We have shown that the BEC of excitonic 
electron-hole pairs cooperatively induces the instability of the phonon 
mode at momentum $q\rightarrow 0$ in the quasi-1D Ta-NiSe-Ta chain, 
resulting in the structural phase transition of the system.  
We have also shown that the spontaneous $c$-$f$ hybridization explains 
the valence states of Ni and Ta ions observed in the XPS experiment 
and that the calculated single-particle spectra reproduce the flattening 
and shift of the band structure observed in the ARPES experiment.  
All of these results support the EI state formation due to the BEC 
mechanism of spin-singlet excitons for the intriguing phase transition 
observed in Ta$_2$NiSe$_5$.  

\begin{acknowledgments}
Enlightening discussions with S. Ejima, H. Fehske, T. Mizokawa, 
K. Seki, Y. Wakisaka, and B. Zenker are gratefully acknowledged.  
This work was supported in part by a Kakenhi Grant 
No.~22540363 of Japan.  A part of computations was carried 
out at the Research Center for Computational Science, 
Okazaki Research Facilities, Japan.  
\end{acknowledgments}


\begin{thebibliography}{99}
%
\bibitem{mott} N. F. Mott, Philos. Mag. \textbf{6}, 287 (1961). 
%
\bibitem{knox} R. Knox, in \textit{Solid State Physics}, edited 
by F. Seitz and D. Turnbull (Academic Press, New York, 1963), 
Suppl. 5, p. 100.  
%
\bibitem{jerome} D. J\'erome, T. M. Rice, and W. Kohn, 
Phys. Rev. \textbf{158}, 462 (1967).  
%
\bibitem{halperin} B. I. Halperin and T. M. Rice, 
Rev. Mod. Phys. \textbf{40}, 755 (1968).  
%
\bibitem{bronold} F. X. Bronold and H. Fehske, 
Phys. Rev. B \textbf{74}, 165107 (2006).  
%
\bibitem{ihle} D. Ihle, M. Pfafferott, E. Burovski, F. X. Bronold, and H. Fehske, 
Phys. Rev. B \textbf{78}, 193103 (2008).  
%
\bibitem{phan} V.-N. Phan, K. W. Becker, and H. Fehske, Phys. Rev. B \textbf{81}, 205117 (2010).  
%
\bibitem{seki} K. Seki, R. Eder, and Y. Ohta, 
Phys. Rev. B \textbf{84}, 245106 (2011).  
%
\bibitem{zenker} B. Zenker, D. Ihle, F. X. Bronold, and H. Fehske, 
Phys. Rev. B \textbf{85}, 121102(R) (2012).  
%
\bibitem{bucher} B. Bucher, P. Steiner, and P. Wachter, 
Phys. Rev. Lett. \textbf{67}, 2717 (1991).  
%
\bibitem{wachter1} P. Wachter, Solid State Commun. \textbf{118}, 645 (2001). 
%
\bibitem{wachter2} P. Wachter, B. Bucher, and J. Malar, 
Phys. Rev. B \textbf{69}, 094502 (2004).  
%
\bibitem{bronold2} F. X. Bronold, H. Fehske, and G. R\"opke, 
J. Phys. Soc. Jpn. \text{76}, Suppl. A, 27 (2007).  
%
\bibitem{young} D. P. Young, D. Hall, M. E. Torelli, Z. Fisk, 
J. L. Sarrao, J. D. Thompson, H.-R. Ott, S. B. Oseroff, 
R. G. Goodrich, and R. Zysler, Nature \textbf{397}, 412 (1999).  
%
\bibitem{zhitomirsky} M. E. Zhitomirsky, T. M. Rice, and 
V. I. Anisimov, Nature \textbf{402}, 251 (1999).  
%
\bibitem{bascones} E. Bascones, A. A. Burkov, and A. H. MacDonald, 
Phys. Rev. Lett. \textbf{89}, 086401 (2002).  
%
\bibitem{cercellier} H. Cercellier, C. Monney, F. Clerc, C. Battaglia, 
L. Despont, M. G. Garnier, H. Beck, P. Aebi, L. Patthey, H. Berger, 
and L. Forr\'o, Phys. Rev. Lett. \textbf{99}, 146403 (2007).    
%
\bibitem{wezel} J. van Wezel, P. Nahai-Williamson, and S. S. Saxena, 
Phys. Rev. B \textbf{81}, 165109 (2010).  
%
\bibitem{monney} C. Monney, C. Battaglia, H. Cercellier, P. Aebi, and H. Beck, 
Phys. Rev. Lett. \textbf{106}, 106404 (2011).  
%
\bibitem{brydon} P. M. R. Brydon and C. Timm, Phys. Rev. B \textbf{80}, 174401 (2009).  
%
\bibitem{zocher} B. Zocher, C. Timm, and P. M. R. Brydon, Phys. Rev. B \textbf{84}, 144425 (2011).  
%
\bibitem{kaneko} T. Kaneko, K. Seki, and Y. Ohta, Phys. Rev. B \textbf{85}, 165135 (2012).  
%
\bibitem{eisenstein} J. P. Eisenstein and A. H. MacDonald, Nature \textbf{432}, 691 (2004).  
%
\bibitem{wakisaka1} Y. Wakisaka, T. Sudayama, K. Takubo, T. Mizokawa, 
M. Arita, H. Namatame, M. Taniguchi, N. Katayama, M. Nohara, and H. Takagi, 
Phys. Rev. Lett. \textbf{103}, 026402 (2009) 
%
\bibitem{wakisaka2} Y. Wakisaka, T. Sudayama, K. Takubo, T. Mizokawa, N. L. Saini, 
M. Arita, H. Namatame, M. Taniguchi, N. Katayama, M. Nohara, and H. Takagi, 
J. Supercond. Nov. Magn. \textbf{25}, 1231 (2012).  
%
\bibitem{sunshine} S. A. Sunshine and J. A. Ibers, 
Inorg. Chem. \textbf{24}, 3611 (1985).  
%
\bibitem{disalvo} F. J. DiSalvo, C. H. Chen, R. M. Fleming, 
J. V. Waszczak, R. G. Dunn, S. A. Sunshine, and J. A. Ibers, 
J. Less-Common Met. \textbf{116}, 51 (1986).  
%
\bibitem{canadell} E. Canadell and M.-H. Whangbo, 
Inorg. Chem. \textbf{26}, 3974 (1987).  
%
\bibitem{wien2k} P. Blaha, K. Schwarz, G. K. H. Madsen, 
D. Kvasnicka, and J. Luitz, \textit{WIEN2K} 
(Technische Universit\"at Wien, Austria, 2002).  
%
\bibitem{PBE96} J. P. Perdew, K. Burke, and M. Ernzerhof, 
Phys. Rev. Lett. \textbf{77}, 3865 (1996).  
%
\bibitem{SM} See Supplemental Material for details of the band structure 
calculations and mean-field equations.  
%
\bibitem{park} S.-G. Park, B. Magyari-K\"ope, and Y. Nishi, 
Phys. Rev. B \textbf{82}, 115109 (2010).  
%
\bibitem{lin} H. Lin, R. S. Markiewicz, L. A. Wray, L. Fu, M. Z. Hasan, and A. Bansil,
Phys. Rev. Lett. \textbf{105}, 036404 (2010); also see e-print arXiv:1003.2615.  
%
\bibitem{ssh} W. P. Su, J. R. Schrieffer, and A. J. Heeger, 
Phys. Rev. B \textbf{22}, 2099 (1980); \textbf{28}, 1138 (1983).  
%
\end{thebibliography}
\end{document}